\documentclass{article}

\usepackage{amssymb,amsfonts,amsmath,stmaryrd}
\usepackage{cite,enumerate,float,indentfirst}
\usepackage{color}
\usepackage{tikz}
\usetikzlibrary{arrows,snakes,backgrounds}
\usepackage{ytableau}
\usepackage[vcentermath]{youngtab}
\usepackage{multicol}
\usepackage{url}

\def\be{\begin{eqnarray}}
\def\ee{\end{eqnarray}}
\def\nn{\nonumber}

\def\Tr{{\rm Tr}\,}

\def\qD{\hbox{qD}}
\def\Ad{\widehat{\rm\bf Ad}}

\definecolor{red}{rgb}{1,0,0}
\definecolor{orange}{rgb}{1,0.5,0}
\definecolor{violet}{rgb}{0.7,0,1}

\hoffset= - 1.0in         

\begin{document}
\title{\vspace{1.5cm}\bf\Large
Torus knots in adjoint representation and Vogel's universality
}

\author{
Liudmila Bishler$^{a,b,}$\footnote{bishlerlv@lebedev.ru},
Andrei Mironov$^{a,b,c,}$\footnote{mironov@lpi.ru,mironov@itep.ru}
}

\date{ }

\maketitle

\vspace{-6cm}

\begin{center}
  \hfill FIAN/TD-07/25\\
  \hfill ITEP/TH-20/25\\
   \hfill IITP/TH-17/25
\end{center}

\vspace{4cm}

\begin{center}
$^a$ {\small {\it Lebedev Physics Institute, Moscow 119991, Russia}}\\
$^b$ {\small {\it Institute for Information Transmission Problems, Moscow 127994, Russia}}\\
$^c$ {\small {\it NRC ``Kurchatov Institute", 123182, Moscow, Russia}}
\end{center}

\vspace{.1cm}

\begin{abstract}
Vogel's universality gives a unified description of the adjoint  sector of representation theory
for simple Lie algebras in terms of three parameters $\alpha,\beta,\gamma$, which are homogeneous coordinates of Vogel's plane.
It is associated with representation theory within the framework of Chern-Simons theory only, and gives rise to universal knot invariants. We extend the list of these latter further, and explain how to deal with the adjoint invariants for the torus knots $T[m,n]$ considering the case of $T[4,n]$ with odd $n$ in detail.
\end{abstract}

\bigskip

\newcommand\smallpar[1]{
  \noindent $\bullet$ \textbf{#1}
}

\section{Introduction}

Three decades ago, P. Vogel discovered a universality \cite{Vogel95,Vogel99} (see also a recent review in \cite{KMS}): this is the claim that the simple Lie algebras are associated
with some isolated points at three lines in the Vogel's projective plane parameterized by three parameters\footnote{One can scale all of these parameters at once with an arbitrary constant. One usually chooses one of the parameters, $\mathfrak{a}$ to be -2. Note that these parameters are usually denoted as $\alpha$, $\beta$ and $\gamma$.} $\mathfrak{a}$, $\mathfrak{b}$, $\mathfrak{c}$, and there are universal algebraic quantities, which are {\bf symmetric functions} of these parameters. These universal quantities are: the Chern-Simons partition function \cite{MkrtVes12,Mkrt13,KreflMkrt,M2}, the dimension \cite{Vogel99} and quantum dimension \cite{Westbury03,MkrtQDims} of the adjoint representation, eigenvalues of the second and higher Casimir operators \cite{LandMan06,MkrtSergVes,ManeIsaevKrivMkrt, IsaevProv, IsaevKriv,IsaevKrivProv} in these representations, the volume of simple Lie groups \cite{KhM},
the HOMFLY-PT knot/link polynomial colored with adjoint representation \cite{MMMuniv,MMuniv} and the Racah matrix involving the adjoint representation and its descendants \cite{MMuniv,ManeIsaevKrivMkrt, IsaevProv, IsaevKriv,IsaevKrivProv} (see also \cite{KLS}).

The Vogel's parameters for simple Lie algebras are listed in Table \ref{vogelparm}. We will mostly use the parameters $u:=q^\mathfrak{a}$, $v:=q^\mathfrak{b}$, $w:=q^\mathfrak{c}$ and $T:=q^\mathfrak{t}=uvw$.
\begin{table}[!ht]
\centering
\begin{tabular}{|c|c|c|c|c|c|}
\hline
Root system & Lie algebra & $\mathfrak{a}$ & $\mathfrak{b}$ & $\mathfrak{c}$ & $\mathfrak{t} = \mathfrak{a}+\mathfrak{b}+\mathfrak{c}$ \\
\hline
$A_n$ & ${sl}_{n+1}$ & $-2$ & $2$ & $n+1$ & $n+1$ \\
$B_n$ & ${so}_{2n+1}$ & $-2$ & $4$ & $2n-3$ & $2n-1$ \\
$C_n$ & ${sp}_{2n}$ & $-2$ & $1$ & $n+2$ & $n+1$ \\
$D_n$ & ${so}_{2n}$ & $-2$ & $4$ & $2n-4$ & $2n-2$ \\
$G_2$ & ${g}_2$ & $-2$ & $\frac{10}{3}$ & $\frac{8}{3}$ & $4$ \\
$F_4$ & ${f}_4$ & $-2$ & $5$ & $6$ & $9$ \\
$E_6$ & ${e}_6$ & $-2$ & $6$ & $8$ & $12$ \\
$E_7$ & ${e}_7$ & $-2$ & $8$ & $12$ & $18$ \\
$E_8$ & ${e}_8$ & $-2$ & $12$ & $20$ & $30$ \\
\hline
\end{tabular}
\caption{Vogel's parameters}
\label{vogelparm}
\end{table}

Note that the Vogel's universality is rather associated not with representation theory of algebras but with Chern-Simons/knot theory: all the universal quantities are this or that way related to Wilson averages in this theory (knot invariants). This is not that much surprising since P. Vogel originally obtained his universality from knot theory. This point becomes especially transparent after the Macdonald deformation, or coming to the refined Chern-Simons theory \cite{BMM}.

Earlier, there were presented manifest constructions for universal knot and link invariants in the adjoint representation in the case of torus knots and links $T[2,n]$ and $T[3,n]$ \cite{MMMuniv} and for twisted knots \cite{MMuniv}.

The next important step is to make it for other torus cases, in particular, for the torus case $T[4,n]$: first of all, it has various applications \cite{Vivek} and, second, it allows one to further study the structure of universality. In fact, everything needed to evaluate the universal invariant of the torus links $T[4,4n]$ in the adjoint representation has been already constructed in \cite{ManeIsaevKrivMkrt}. Indeed, this invariant is given by the Rosso-Jones formula for links \cite{RJ,China1,Stevan,China2}:
\be
P^{[4,4n]}_{Adj}(q)\ = \ \frac{  q^{16n\varkappa_R}}{\qD_{\!_R}(q )} \sum_{Q\in R^{\otimes 4}}
N_Q \cdot q^{-n\varkappa_{_Q}}\cdot \qD_{_Q}(q )
\label{RJl}
\ee
where $\varkappa_Q$ is the eigenvalue of the second Casimir operator, $\qD_Q$ is the quantum dimension, and $N_Q$ is the number of times the irrep $Q$ is met in the decomposition of the fourth power of the adjoint representation, $Adj^{\otimes 4}$. The most non-trivial part of this formula is just these coefficients $N_Q$, and they are calculated in \cite{ManeIsaevKrivMkrt}.

However, evaluating the adjoint link invariant in the case of the torus link $T[4,4n]$ though immediate still requires rather massive calculations: the sum in (\ref{RJl}) contains a lot of terms (49 terms), and one still has to construct universal quantum dimensions for each of them. Note that many of them are not factorized being sums of irreps (called in \cite{BMM} uirreps) for concrete algebras: they are have just the same eigenvalues of the second Casimir operator, and are called in \cite{ManeIsaevKrivMkrt} Casimir eigenspaces. In fact, the Casimir eigenspaces are exactly what one needs when constructing knot/link invariants: contributions of irreps from the same Casimir eigenspace are merely summed: the common coefficient is just $q^{\varkappa_Q}$ \cite{modRT}.

At the same time, calculation of the adjoint knot invariant in the case of the torus knot $T[4,n]$ (i.e. in the case of odd $n$) is simpler, since the sum similar to (\ref{RJl}) contains in this case only 15 terms in the generic case: the coefficients $N_Q$ in the sum are replaced by the so-called Adams coefficients, which may be both positive and negative. We explain how to find these coefficients for all simple Lie algebras, and evaluate the universal invariant $P^{[4,n]}_{Adj}(q)$ calculating all necessary quantum dimensions both for the concrete Lie algebras and in the universal form. We also discuss the general properties of the universal adjoint polynomials.

\paragraph{Notation.}
Throughout the paper, we use the notation
$$
\{x\}=x-{1\over x}
$$

We denote through $S_R\{p_k\}$ the Schur functions, which are symmetric polynomials of variables $x_i$, or are graded polynomials of the power sums $p_k:=\sum_ix_i^k$. When using the Schur functions for a realization of characters of the representation $R$ of a Lie group, the variables $p_k=\Tr g^k$, where $g$ is the group element, and the trace is evaluated in the defining representation. The Schur functions are labelled by the Young diagrams (partitions), $R=(R_1,R_2,\ldots,R_{l_R})$, $R_1\ge R_2\ge\ldots\ge R_{l_R}>0$, $|R|:=\sum_iR_i$. We also denote through $S_{R/Q}$ the skew Schur functions.

In what follows, we use the notation $T[4,n]$ only for torus knots (unless explicitly stated otherwise), i.e. implying that $n$ is always odd.

For all necessary information about the symmetric functions that are characters of the classical Lie groups, see \cite{SF} and references therein (especially \cite{M,L} and the more modern \cite{BC}), and specifically about the Schur functions, \cite{Mac}.

\section{Rosso-Jones formula}

We start with the Rosso-Jones formula in the universal form for the torus knot $T[m,n]$ ($m$ and $n$ are coprime)
\be
P^{[m,n]}_R(q)\ = \ \frac{  q^{mn\varkappa_R}}{\qD_{\!_R}(q )} \sum_{Q\in R^{\otimes m}}
c_{_{RQ}}^{(m)} \cdot q^{-\frac{n}{m}\varkappa_{_Q}}\cdot \qD_{_Q}(q )
\label{RJ}
\ee
Here
\be
\varkappa_Q = (\Lambda_Q,\Lambda_Q+2\rho)
\ee
is the eigenvalue of the second Casimir operator,
\be\label{Weyl}
D_Q=\prod_{\alpha\in\Delta_+}{[\left(\Lambda_Q+\rho,\alpha\right)]\over [\left(\rho,\alpha\right)]}
\ee
is the quantum dimension, $\Delta_+$ denotes the set of positive roots, and the coefficients $c_{_{RQ}}^{(m)}$ are defined by the Adams operation: $m$-plethysm
\be
\Ad_m \chi_{_R}(p_k) \equiv \chi_{_R}(p_{mk}) = \sum_{Q\in R^{\otimes m}}
c_{_{RQ}}^{(m)} \chi_{_Q}(p_k)
\label{Adrule}
\ee
where $\chi_{_R}(p_{mk})$ is the character of representation $Q$.

Note that one can equivalently obtain the coefficients $c^{(m)}_{RQ}$ in the Adams operation in the following way. Split the decomposition of $R^{\otimes m}$ into a sum of terms with fixed symmetric patterns given by the Young diagrams $P$ of size $m$:
\be
R^{\otimes m}=\sum_{P\vdash m}\pi_P\left(R^{\otimes m}\right)
\ee
Then, the Adams operation gives \cite{RJ,China1}
\be\label{Ad}
\Ad_m (R)=\sum_{P\vdash m}\psi_P([m])\pi_P\left(R^{\otimes m}\right)
\ee
where $\psi_P([m])$ is the value of the character of the permutation group ${\cal S}_m$ in the representation $P$ on the cycle of the maximal length (i.e. on the cyclic permutation $(1,\ldots,m)$).

Note that expression (\ref{RJ}) is symmetric in $m$ and $n$ (which is absolutely non-trivial). The knot invariant (\ref{RJ}) is in the topological framing. Note also that the number of representations $Q$ that contribute to the r.h.s. of the Rosso-Jones formula in this case is much less than in the case of links (\ref{RJl}) (in the case of $m=4$, it is 15 instead of 49, see (\ref{main}) below).

\section{Rosso-Jones formula for classical Lie algebras}

In this section, we apply the general Rosso-Jones formula (\ref{RJ}) to the classical Lie algebras in order to obtain very explicit expressions for the adjoint invariants.

\subsection{$A$ series}

In the $A_{N-1}$ series case, all the ingredients of the Rosso-Jones formula look as follows.

The character is given by the Schur function,
\be
\chi^{A_n}_Q=S_R\{p_m\}
\ee
where $R$ is the Young diagram (partition) labelling the representation.

In particular, the character of the adjoint representation is given by the Schur function $S_{[21^{N-2}]}$, and we introduce the notation $A:=q^N$. Note that this notation is not the second variable $A$ of the HOMFLY-PT polynomials: the HOMFLY-PT polynomial is evaluated for one and the same Young diagram for various $A_{N-1}$, while the adjoint representation depends on $N$ itself. This kind of knot invariants is called uniform \cite{MMMuniv} or composite \cite{ChE}.

The Casimir eigenvalue in this case is
\be
\varkappa_Q = 2\sum_{\Box_{i,j}\in Q}(j-i)-{|Q|^2\over N}+|Q|N
\ee

The 4-plethystic expansion of the adjoint Schur function generating the Adams coefficients is
\be\label{A4}
\Ad_4 S_{Adj}&=&3+S_{[2^41^{N-8}]}-S_{[3^42^{N-7}1^2]}+S_{[4^43^{N-6}2]}-S_{[5^44^{N-5}]}-\nn\\
&-&S_{[32^21^{N-7}]}+S_{[43^22^{N-6}1^2]}-S_{[54^23^{N-5}2]}+S_{[65^24^{N-4}]}+\nn\\
&+&S_{[421^{N-6}]}-S_{[532^{N-5}1^2]}+S_{[643^{N-4}2]}-S_{[754^{N-3}]}-\nn\\
&-&S_{[51^{N-5}]}+S_{[62^{N-4}1^2]}-S_{[73^{N-3}2]}+S_{[84^{N-2}]}
\ee
The simplest technical way to evaluate the Adams coefficients is to use the explicit formula
\be
c_{_{RQ}}^{(m)}=\sum{\psi_R(\Delta)\psi_Q(m\Delta)\over z_\Delta}
\ee
where $\psi_R(\Delta)$ is the character of the representation $R$ of the permutation group on the conjugacy class given by the Young diagram $\Delta$, $z_\Delta$ is the order of automorphism of the Young diagram $\Delta$: if $\Delta=[\ldots,3^{r_3},2^{r_2},1^{r_1}]$, where some $r_k$ may be equal to zero, then $z_\Delta:=\prod_k k^{r_k}r_k!$. The Young diagram $m\Delta$ is understood as the Young diagram with all lengths of lines multiplied by $m$.

The adjoint quantum dimension and Casimir exponential are
\be\label{qDA}
\begin{array}{rclrcl}
\qD_{Adj}=\qD_{[21^{N-2}]}&=&\displaystyle{\{Aq\}\{A/q\}\over\{q\}^2}
\hspace{1cm}&q^{\varkappa_{Adj}} &=&A^2\cr
\end{array}
\ee
and all other quantum dimensions and second Casimir eigenvalues can be found in the Appendix.

Now one can obtain the uniform HOMFLY-PT polynomial $H^{(m,n)}_{Adj}$ of the torus $T[4,n]$ knot using the Rosso-Jones formula (\ref{RJ}) and substituting in it the manifest expressions for $\varkappa_{_Q}$ and $\qD_{_Q}$ from the Appendix. One can easily check that (unknot case)
\be
H^{[4,1]}=1
\ee
and that (pure plethysm)\footnote{The case of $n=0$ does not describe any knot.}
\be
H^{[4,0]}={\qD_{Adj}\Big|_{q\to q^4,A\to A^4}\over\qD_{Adj}}
\ee

\subsection{Orthogonal series}

In the case of other classical algebras the calculation looks simpler, since the adjoint representation is associated with a fixed diagram, with [1,1] for the orthogonal $B_n$ and $D_n$ systems and with [2] for the symplectic $C_n$ case. In the orthogonal case $so(N)$, the character associated with the diagram $Q$ is given by the formula
\be\label{orth}
\chi^{B/D_n}_Q:=So_Q\{p_k\}=\sum_{R}(-1)^{|R|/2}S_{Q/R}\{p_k\}
\ee
where $R$ ranges over the Frobenius coordinates \cite{Mac} $(r_1+1,r_2+1,\ldots|r_1,r_2,\ldots)$ including the empty partition $R=\emptyset$.

Coming to the ingredients of the Rosso-Jones formula, which gives rise to the Kauffman polynomials in this case, we parameterize $A=q^{N/2-1}$ (i.e. $A = q^{n-1/2}$ in the $B_n=so(2n+1)$ case, and $A=q^{n-1}$ in the $D_n=so(2n)$ case), and this time it is just the standard parameter of the Kauffman polynomial $K_R(A,q)$.

The Casimir eigenvalue in this case is
\be
 \varkappa_Q = 2|Q|N+\sum_{i=1}^{l_Q} Q_i(Q_i-2i)
\ee
The 4-plethystic expansion of the adjoint Schur function generating the Adams coefficients is
\be\label{D4}
\Ad_4 So_{Adj}&=&2-So_{[1111]}+So_{[211]}-So_{[31]}+So_{[4]}+\nn\\
&+&So_{[1^8]}-So_{[21^6]}+So_{[2222]}+\nn\\
&+&So_{[31^5]}-So_{[3221]}+So_{[332]}-\nn\\
&-&So_{[41111]}+So_{[4211]}-So_{[431]}+So_{[44]}
\ee

The adjoint quantum dimension and Casimir exponential are
\be\label{qDD}
\begin{array}{rclrcl}
\qD_{Adj}=\qD_{[11]}&=&\displaystyle{\{qA\}\{qA^2\}\{A^2/q^2\}\over\{q\}\{q^2\}\{A/q\}}
\hspace{1cm}&q^{\varkappa_{Adj}} &=&A^4\cr
\end{array}
\ee
and all other quantum dimensions and second Casimir eigenvalues can be found in the Appendix.

Now one can obtain the adjoint Kauffman polynomial $K^{(m,n)}_{Adj}$ of the torus $T[4,n]$ knot using the Rosso-Jones formula (\ref{RJ}) and substituting in it the manifest expressions for $\varkappa_{_Q}$ and $\qD_{_Q}$ from the Appendix. One can easily check that (unknot case)
\be
K^{(4,1)}=1
\ee
and that (pure plethysm)
\be
K^{(4,0)}={\qD_{Adj}\Big|_{q\to q^4,A\to A^4}\over\qD_{Adj}}
\ee

\subsection{Symplectic series}

In the case of symplectic algebra $C_n=sp_{2n}$, the adjoint representation is associated with diagram [2]. In this case, the character associated with the diagram $Q$ is given by the formula
\be\label{sym}
\chi^{C_n}_Q:=Sp_Q\{p_k\}=\sum_{R}(-1)^{|R|/2}S_{Q/R}\{p_k\}
\ee
where $R$ ranges over the Frobenius coordinates \cite{Mac} $(r_1,r_2,\ldots|r_1+1,r_2+1,\ldots)$ including the empty partition $R=\emptyset$.

The symplectic characters are related \cite{R81,Tsv,MV10} with the orthogonal ones by a simple relation $\omega(\chi^{C_n}_Q)=\chi^{D_n}_{Q^\vee}$, where $Q^\vee$ denotes the transposed Young diagram, and $\omega$ is the standard involution on the symmetric group.

To put it differently, since
\be
S_{Q/R}\{p_k\}=(-1)^{|Q|+|R|}S_{Q^\vee/R^\vee}\{-p_k\}
\ee
one immediately obtains from (\ref{orth}) and (\ref{sym}) that
\be
\chi^{C_n}_Q\{p_m\}=\sum_{R}(-1)^{|R|/2}S_{Q/R}\{p_m\}=\sum_{R}(-1)^{|R|/2+|R|+|Q|}S_{Q^\vee/R^\vee}\{-p_m\}=(-1)^{|Q|}\chi^{B/D_n}_{Q^\vee}\{-p_m\}
\ee
since $|R|$ is even.

This gives rise, instead of (\ref{D4}), to the following Adams operation
\be
\label{C4}
\Ad_4 Sp_{Adj}&=&2+Sp_{[1111]}-Sp_{[211]}+Sp_{[31]}-Sp_{[4]}+\nn\\
&+&Sp_{[2222]}-Sp_{[3221]}+Sp_{[332]}+\nn\\
&+&Sp_{[4211]}-Sp_{[431]}+Sp_{[44]}-\nn\\
&-&Sp_{[5111]}+Sp_{[611]}-Sp_{[71]}+Sp_{[8]}
\ee
It also results to the substitution $q^n\to q^{-n}$ in all quantum dimensions \cite{BFM} and in the Kauffman invariants:
\be
\qD_Q^{C_n}(q^n,q)=(-1)^{|Q|}\qD^{D_n}_{Q^\vee}(q^{-n},q)
\ee
Hence, they can be read immediately from the orthogonal formulas, and we do not write them down here.

\section{Universal Adams operation for $T[4,n]$}

\subsection{The universal structure}

In order to obtain the universal formula for the adjoint polynomial, we use the universal decomposition of the fourth power of the adjoint representation in \cite{ManeIsaevKrivMkrt}. One can apply formula (\ref{Ad}0 to the torus knot $T[4,n]$, using the decomposition of $Adj^{\otimes 4}$ in \cite{ManeIsaevKrivMkrt}, and the values
\be
\psi_{[4]}([4])=1,\ \ \ \ \ \psi_{[31]}([4])=-1,\ \ \ \ \ \psi_{[22]}([4])=0,\ \ \ \ \ \psi_{[211]}([4])=1,\ \ \ \ \ \psi_{[1111]}([4])=-1
\ee
It results into the Adams operation
\be
\Ad_4(Adj)=2+ X_2-\mathbb{X}_3-X_4+J+J'+J''+\mathbb{Z}_3+Y_4+Y'_4+Y''_4-\mathbb{K}_3-G-G'-G''+\mathbb{L}_3+I+I'+I''
\ee
where we used the notation from \cite{ManeIsaevKrivMkrt}. In particular, $J':=J\Big|_{\mathfrak{a}\leftrightarrow\mathfrak{b}}=J\Big|_{u\leftrightarrow v}$; $J'':=J\Big|_{\mathfrak{a}\leftrightarrow\mathfrak{c}}=J\Big|_{u\leftrightarrow w}$, etc. Note that, in the $A$ series case,
\be
\mathbb{Z}_3&=&2\hat X_3\nn\\
\mathbb{X}_3&=&\hat X_3+\tilde X_3\nn\\
\mathbb{K}_3&=&3\hat X_3+\tilde X_3\nn\\
\mathbb{L}_3&=&2\hat X_3+2\tilde X_3
\ee
while, for the other simple Lie algebras,
\be
\mathbb{Z}_3=\mathbb{X}_3\nn\\
\mathbb{K}_3=\mathbb{L}_3
\ee
Thus, the combination
\be\label{open}
-\mathbb{X}_3+\mathbb{Z}_3-\mathbb{K}_3+\mathbb{L}_3=0
\ee
always vanishes, and we finally obtain:
\be\label{main}
\boxed{
\Ad_4(Adj)=2+ X_2-X_4+J+J'+J''+Y_4+Y'_4+Y''_4-G-G'-G''+I+I'+I''
}
\ee

\subsection{$A$ series}

Consider how this formula works in the $A$ series case.
Formulas (\ref{qDA}) imply that the irreps emerging after the Adams operation are divided into ten groups of the Casimir eigenspaces, which we called uirreps in \cite{BMM}. The irreps from the same group have the same eigenvalues of the second Casimir operator, the same dimensions and the same quantum dimensions:
\be\label{A4c}
Q_1=[2^41^{N-8}]=Y_4'\nn\\
Q_2=[3^42^{N-7}1^2]\oplus [3221^{N-7}]=G'\nn\\
Q_3=[4^43^{N-6}2]\oplus[421^{N-6}]\nn\\
Q_4=[5^44^{N-5}]\oplus [51^{N-5}]\nn\\
Q_5=[4332^{N-6}1^2]\nn\\
Q_6=[5443^{N-5}2]\oplus [532^{N-5}1^2]\nn\\
Q_7=[6554^{N-4}]\oplus [62^{N-4}11]\nn\\
Q_8=[643^{N-4}2]\nn\\
Q_9=[754^{N-3}]\oplus [73^{N-3}2]=G\nn\\
Q_{10}=[84^{N-2}]=Y_4
\ee
Since, in this case,
\be
Y_4''=0\nn\\
J=0\nn\\
J'=0\nn\\
G''=-1
\ee
and
\be
X_2+I''=0\nn\\
I'=Q_3+Q_5\nn\\
I=Q_7+Q_8\nn\\
X_4-J''=Q_4+Q_6
\ee
we finally obtain from (\ref{main}) the Adams operation
\be
\Ad_4(Adj)=3+Q_1-Q_2+Q_3-Q_4+Q_5-Q_6+Q_7+Q_8-Q_9+Q_{10}
\ee
which coincides with (\ref{A4}).

In order to illustrate how these formulas work in terms of irreps, we note that $X_4=Q_4+Q_6+[4^2,2^{N-4}]$, while\footnote{This representation is characterized by
$$
\begin{array}{rclrcl}
\qD_{[4^2,2^{N-4}]}&=&\displaystyle{\{A/q^3\}\{A/q^2\}^2\{A/q\}\{Aq\}\{Aq^2\}^2\{Aq^3\}\over\{q\}^2\{q^2\}^4\{q^3\}^2}
\hspace{1cm}&q^{\varkappa_{[4^2,2^{N-4}]}}&=&A^8
\end{array}
$$
}
$J''=[4^2,2^{N-4}]$. Note that $[4^2,2^{N-4}]$ is the irrep that does not appear in the decomposition (\ref{A4}), but it is necessary for restoring the universal form. However, all combinations of this kind have the same second Casimir operator eigenvalue, since the decomposition (\ref{main}) is into the Casimir eigenspaces. This means that all of them have the same factor $q^{-\frac{n}{m}\varkappa_{_Q}}$ in the Rosso-Jones formula, and the decomposition does not change with changing $n$.

\subsection{On phantom (virtual) representations}

Note that in the formulas of the previous subsection, there are relations $X_2+I''=0$ and $G''+1=0$, which imply that some representations are negative. It is certainly just a notation, which means the following: these are ordinary irreps for some values of the Vogel's parameters, but at other values of parameters they are no longer representations at all, and, moreover, have formally negative dimensions (and quantum dimensions with a ``wrong" sign). This is the consequence of the fact that the Vogel's universality is associated not with representation theory but with Chern-Simons/knot theory. In knot theory, one works only with the (quantum) dimensions of representations, and the negative dimension of the representation is formally allowed, when in the formula for knot invariant it enters with the negative sign. In the examples above, we see that the coefficient of the singlet contribution to the adjoint invariant, i.e. in front of zero Casimir eigenvalue term in (\ref{A4}) is equal to 3, while the universal formula (\ref{main}) gives only the coefficient 2. However, in the universal formula for the invariant, there is also a contribution -1 coming from the $G''$ term (which is also associated with the zero Casimir eigenvalue) at the values of $u=q^{-2}$, $v=q^2$, $w=A$ describing the $A$ series. This is what we denote by the sign minus in front of representation of the particular algebra.

In such cases, this kind of contributions is called ``phantom" or ``virtual" representations \cite{ManeIsaevKrivMkrt}, since these are not representations of the particular algebra but just a technical trick to describe invariants in a universal way for all simple algebras at once.

In fact, as we shall see, $I''$ is a phantom representation for any simple algebra. However, it is necessary to have formulas symmetric in the Vogel's parameters, on one hand, and is necessary to reproduce proper negative contributions in the universal adjoint invariant, on the other hand. A similar situation is with representations $G''$, $J$ and $Y_4''$: for any concrete simple algebra, they are either zero (i.e. do not contribute), or are phantom representations. At the same time, representations $Y_4'$ and $G'$ become phantom only for the exceptional algebras.

\subsection{Orthogonal series}

In this case, there are minimal number irreps at the r.h.s. of (\ref{main}) that do not contribute: just two
\be
Y''_4=0\nn\\
G''=0
\ee
On the other hand, remaining correspondences are simpler, and are as follows:
\be\label{D4c}
\phantom{.}[211]&=&X_2\nn\\
\phantom{.}[3221]\oplus[41^4]&=&X_4\nn\\
\phantom{.}[1111]&=&-J\nn\\
\phantom{.}[4]&=&J'\nn\\
\phantom{.}[2222]&=&J''\nn\\
\phantom{.}[44]&=&Y_4\nn\\
\phantom{.}[1^8]&=&Y_4'\nn\\
\phantom{.}[431]&=&G\nn\\
\phantom{.}[21^6]&=&G'\nn\\
\phantom{.}[332]\oplus[4211]&=&I\nn\\
\phantom{.}[31^5]&=&I'\nn\\
\phantom{.}[31]&=&-I''
\ee
With these correspondences, formula (\ref{main}) gives rise to (\ref{D4}). Note that, in variance with the $A$ series case (\ref{A4c}), the irreps in (\ref{D4c}) entering the same Casimir eigenspaces have {\bf distinct} dimensions (but certainly have coinciding Casimir eigenvalues). Here there are two phantom representations: $J$ and again $I''$.

\subsection{Exceptional series}

In the case of exceptional algebras, as usual, there are more irreps that do not appear in the decomposition (\ref{main}). In particular,
\be
J=0\nn\\
J'=0\nn\\
G''=0\nn\\
I'=0\nn\\
Y''_4=-1
\ee
Hence, one again remains with just 10 terms in the Adams operation:
\be
\Ad_4(Adj)=1+ X_2-X_4+J''+Y_4+Y'_4-G-G'+I+I''
\ee
The full list of the remaining representations (with their ordinary dimensions) in this case looks as follows:

\bigskip

\framebox[1cm]{\bf E6}
\begin{equation}
    \begin{aligned}
   Adj = \textbf{78} & \quad\quad    \omega_{Adj} = \omega_6 \\
    X_2 = \textbf{2925} & \quad\quad   \omega_{X_2} = \omega_3   \\
    X_4 = \textbf{600600}\oplus\overline{\textbf{600600}} & \quad\quad   \omega_{X_4} = \omega_1+2\omega_4,\,\, \omega'_{X_4} = 2\omega_2+\omega_5   \\ 
     J'' = \textbf{85293} & \quad\quad   \omega_{J''} = 2\omega_1+2\omega_5   \\
    Y_4 = \textbf{537966} & \quad\quad     \omega_{Y_4} = 4\omega_6\\
    Y_4' = \textbf{78} = Adj & \quad\quad     \omega_{Y_4'} = \omega_6 = \omega_{Adj}\\
    G = \textbf{1911195} & \quad\quad     \omega_{G} = \omega_3+2\omega_6\\
    G' = \textbf{2925} = X_2 & \quad\quad     \omega_{G'} = \omega_3 = \omega_{X_2}\\
    I = \textbf{2453814} & \quad\quad     \omega_{I} = \omega_2+\omega_4+\omega_6\\
    I'' = \textbf{34749} & \quad\quad     \omega_{I''} = \omega_1+\omega_5+\omega_6\\
    \end{aligned}
\end{equation}
\vspace{10pt}

\noindent
\begin{minipage}[t]{0.48\textwidth}
\framebox[1cm]{\bf E7}
\begin{equation*}
    \begin{aligned}
    Adj = \textbf{133} & \quad\quad    \omega_{Adj} = \omega_1 \\
    X_2 = \textbf{8645} & \quad\quad   \omega_{X_2} = \omega_2   \\
    X_4 = \textbf{11316305} & \quad\quad   \omega_{X_4} = \omega_4+\omega_7\\
     J'' = \textbf{617253} & \quad\quad   \omega_{J''} = 2\omega_5   \\
    Y_4 = \textbf{5248750} & \quad\quad     \omega_{Y_4} = 4\omega_1\\
    Y_4' = \textbf{0} & \quad\quad    \\
    G = \textbf{19046664} & \quad\quad     \omega_{G} = 2\omega_1+\omega_2\\
    G' = \textbf{0}  & \quad\quad  \\
    I = \textbf{24386670} & \quad\quad     \omega_{I} = \omega_1+\omega_3\\
    I'' = \textbf{152152} & \quad\quad     \omega_{I''} = \omega_1+\omega_5\\
    \end{aligned}
\end{equation*}
\end{minipage}
\noindent
\begin{minipage}[t]{0.48\textwidth}
\framebox[1cm]{\bf E8}
\begin{equation*}
    \begin{aligned}
  Adj = \textbf{248} & \quad\quad    \omega_{Adj} = \omega_7 \\
    X_2 = \textbf{30380} & \quad\quad   \omega_{X_2} = \omega_6   \\
    X_4 = \textbf{146325270} & \quad\quad   \omega_{X_4} = \omega_4   \\
     J'' = \textbf{4881384} & \quad\quad   \omega_{J''} = 2\omega_1   \\
    Y_4 = \textbf{79143000} & \quad\quad     \omega_{Y_4} = 4\omega_7\\
    Y_4' = \textbf{3875} & \quad\quad     \omega_{Y_4'} = \omega_1 \\
    G = \textbf{281545875} & \quad\quad     \omega_{G} = \omega_6+2\omega_7\\
    G' = \textbf{147250}  & \quad\quad     \omega_{G'} = \omega_8\\
    I = \textbf{344452500} & \quad\quad     \omega_{I} = \omega_5+\omega_7\\
    I'' = \textbf{779247} & \quad\quad     \omega_{I''} = \omega_1+\omega_7\\
    \end{aligned}
\end{equation*}
\end{minipage}
\vspace{10pt}

\noindent
\begin{minipage}[t]{0.48\textwidth}
\framebox[1cm]{\bf F4}
\begin{equation*}
    \begin{aligned}
  Adj = \textbf{52} & \quad\quad    \omega_{Adj} = \omega_1 \\
    X_2 = \textbf{1274} & \quad\quad   \omega_{X_2} = \omega_2   \\
    X_4 = \textbf{205751} & \quad\quad   \omega_{X_4} = 2\omega_3+\omega_4  \\
     J'' = \textbf{16302} & \quad\quad   \omega_{J''} = 4\omega_4  \\
    Y_4 = \textbf{100776} & \quad\quad     \omega_{Y_4} = 4\omega_1\\
    Y_4' = \textbf{26} & \quad\quad     \omega_{Y_4'} = \omega_4 \\
    G = \textbf{340119} & \quad\quad     \omega_{G} = 2\omega_1+\omega_2\\
    G' = \textbf{1053}  & \quad\quad     \omega_{G'} = \omega_1+\omega_4\\
    I = \textbf{420147} & \quad\quad     \omega_{I} = \omega_1+2\omega_3\\
    I'' = \textbf{10829} & \quad\quad     \omega_{I''} = \omega_1+2\omega_4\\
    \end{aligned}
\end{equation*}
\end{minipage}
\begin{minipage}[t]{0.48\textwidth}
\framebox[1cm]{\bf G2}
\begin{equation*}
    \begin{aligned}
  Adj = \textbf{14} & \quad\quad    \omega_{Adj} = \omega_2 \\
    X_2 = \textbf{77} & \quad\quad   \omega_{X_2} = 3\omega_1   \\
    X_4 = \textbf{0} & \quad\quad    \\
     J'' = \textbf{0} & \quad\quad    \\
    Y_4 = \textbf{748} & \quad\quad     \omega_{Y_4} = 4\omega_2\\
    Y_4' = \textbf{0} & \quad\quad   \\
    G = \textbf{1547} & \quad\quad     \omega_{G} = 3\omega_1+2\omega_2\\
    G' = \textbf{0}  & \quad\quad  \\
    I = \textbf{924} &\quad\quad     \omega_{I} = 4\omega_1+\omega_2\\
    I'' = \textbf{189} & \quad\quad     \omega_{I''} = 2\omega_1+\omega_2\\
    \end{aligned}
\end{equation*}
\end{minipage}

\section{Universal adjoint invariant for $T[4,n]$}

\subsection{Universal invariant explicitly}

Now we are ready to construct the universal adjoint invariant. It has the form
\be
\boxed{
\begin{array}{c}
P^{[4,n]}_{Adj}(u,v,w)\ = \ \displaystyle{  T^{8n}\over \qD_{\!_{Adj}}(u,v,w )} \Big(
2+ T^{-n}\cdot\qD_{X_2}-T^{-2n}\cdot\qD_{X_4}+T^{-2n}v^nw^n\cdot\qD_{J}
+T^{-2n}u^nw^n\cdot\qD_{J'}+\\
\\
+T^{-2n}v^nu^n\cdot\qD_{J''}
+T^{-2n}u^{3n}\cdot\qD_{Y_4}+T^{-2n}v^{3n}\cdot\qD_{Y'_4}+T^{-2n}w^{3n}\cdot\qD_{Y''_4}
-T^{-2n}u^{2n}\cdot\qD_{G}-\\
\\
-T^{-2n}v^{2n}\cdot\qD_{G'}-T^{-2n}w^{2n}\cdot\qD_{G''}+T^{-2n}u^{n}\cdot\qD_{I}+T^{-2n}v^{n}\cdot\qD_{I'}+T^{-2n}w^{n}\cdot\qD_{I''}
\Big)
\end{array}
}
\label{RJu}
\ee
where the quantum dimensions\footnote{Part of these quantum dimensions can be found in \cite{AM1902}. In their notation, $Y_4$ is described by $k=0$, $n=4$, and $G$, by $k=1$ and $n=2$. Notice a misprint: in $L_{11s1}$, there should be $\alpha (i+1)$ instead of $\alpha (i+2)$ in the denominator.} and the second Casimir eigenvalues can be found in Table 2.

\bigskip

\begin{table}[h]\label{qDu}
\begin{tabular}{|c||c|c|}
\hline&&\cr
Q&$\qD_Q$&$q^{\varkappa_Q}$\cr&&\cr
\hline
\hline
&&\cr
Adj&$-\displaystyle{\left\{\frac{T}{\sqrt{u}}\right\}\left\{\frac{T}{\sqrt{v}}\right\}\left\{\frac{T}{\sqrt{w}}\right\}\over
    \{\sqrt{u}\}
    \{\sqrt{v}\}\{\sqrt{w}\}}$&$T^2$\cr&&\cr
\hline&&\cr
$X_2$&$-\qD_{Adj}\times\displaystyle{\left\{\sqrt{Tu}\right\}\left\{\sqrt{Tv}\right\}
    \left\{\sqrt{Tw}\right\}\left\{\frac{T}{u}\right\}\left\{\frac{T}{v}\right\}\left\{\frac{T}{w}\right\}\over
    \left\{u\right\}\left\{v\right\}\left\{w\right\}
    \left\{\sqrt{\frac{T}{u}}\,\right\}\left\{\sqrt{\frac{T}{v}}\,\right\}\left\{\sqrt{\frac{T}{w}}\,\right\}}$
&$T^4$\cr&&\cr
\hline&&\cr
$X_4$&Formula (\ref{qDX4})
&$T^8$\cr&&\cr
\hline&&\cr
J&$-\displaystyle{\{\sqrt{T}\}\{T\}\{uv\}\{uw\}\left\{{uv\over\sqrt{w}}\right\}\left\{{uw\over\sqrt{v}}\right\}\left\{{u\sqrt{w\over v}}\right\}
\left\{{u\sqrt{v\over w}}\right\}\{\sqrt{Tv}\}\{\sqrt{Tw}\}
\left\{{T\over\sqrt{u}}\right\}\left\{{T\over\sqrt{v}}\right\}\left\{{T\over\sqrt{w}}\right\}\over
\{\sqrt{u}\}\{u\}\{\sqrt{v}\}\{v\}\{\sqrt{w}\}\{w\}\left\{\sqrt{u\over v}\right\}\left\{\sqrt{u\over w}\right\}\left\{{w\over\sqrt{v}}\right\}
\left\{{v\over\sqrt{w}}\right\}\left\{\sqrt{u\over vw}\right\}\left\{\sqrt{v\over w}\right\}^2}$
&$T^8v^{-4}w^{-4}$\cr&&\cr
\hline
&&\cr
$Y_4$&$\displaystyle{\{T\}\left\{{T\over u^{7/2}}\right\}\left\{{T\over\sqrt{u}}\right\}\left\{{T\over\sqrt{v}}\right\}\left\{{T\over\sqrt{w}}\right\}\left\{{T\over u}\right\}
\left\{{T\over\sqrt{uv}}\right\}\left\{{T\over u\sqrt{v}}\right\}\left\{{T\over u\sqrt{uv}}\right\}
\left\{{T\over\sqrt{uw}}\right\}\left\{{T\over u\sqrt{w}}\right\}\left\{{T\over u\sqrt{uw}}\right\}\over \{u^{1/2}\}\{u\}\{u^{3/2}\}\{u^2\}
\{\sqrt{v}\}\{\sqrt{w}\}
\left\{\sqrt{u\over v}\right\}\left\{u\over \sqrt{v}\right\}\left\{{u\sqrt{u\over v}}\right\}
\left\{\sqrt{u\over w}\right\}\left\{u\over \sqrt{w}\right\}\left\{{u\sqrt{u\over w}}\right\}}$
& $T^8u^{-12}$\cr&&\cr
\hline&&\cr
G&${\{uv\}\{uw\}\left\{{u^2\over vw}\right\}\{v\sqrt{w}\}\{w\sqrt{v}\}\{T\}\{\sqrt{uT}\}\{\sqrt{vT}\}\{\sqrt{wT}\}
\left\{{vw\over\sqrt{u}}\right\}\left\{v\sqrt{w\over u}\right\}\left\{w\sqrt{v\over u}\right\}
\left\{{T\over\sqrt{u}}\right\}\left\{{T\over\sqrt{v}}\right\}\left\{{T\over\sqrt{w}}\right\}\over
\{\sqrt{uv}\}\{\sqrt{uw}\}\left\{{u\over \sqrt{vw}}\right\}\{\sqrt{u}\}^2\{u\}\{u^2\}\{\sqrt{v}\}\{\sqrt{w}\}
\left\{{u\over\sqrt{v}}\right\}\left\{{u\over v}\right\}\left\{{\sqrt{u\over v}}\right\}
\left\{{u\over\sqrt{w}}\right\}\left\{{u\over w}\right\}\left\{{\sqrt{u\over w}}\right\}
}$
&$T^8u^{-8}$\cr&&\cr
\hline&&\cr
I& sec.\ref{qDI}
&$T^8u^{-4}$\cr&&\cr
\hline
\end{tabular}
\caption{Table of the quantum dimensions and the second Casimir eigenvalues entering the universal adjoint invariant.}
\end{table}

Calculating the universal quantum dimensions of $\qD_{X_4}$ and $\qD_{I}$ is more involved, and we described it in the next two subsections.

\subsection{Universal quantum dimension $\qD_{X_4}$}

The simplest way to calculate the universal quantum dimension of $X_4$, one can use that it emerges in the antisymmetric cube and the fourth power of the adjoint representation:
\be
\Lambda^3(Adj)&=&1+X_2+\mathbb{X}_3+Y_2+Y'_2+Y''_2\nn\\
\Lambda^4(Adj)&=&Adj+X_2+\mathbb{X}_3+X_4+C+C'+C''+B+B'+B''+Y_2+Y'_2+Y''_2
\ee
where we again used the notation from \cite{ManeIsaevKrivMkrt}. Thus, one obtains
\be
X_4=1+\Lambda^4(Adj)-\Lambda^3(Adj)-Adj-C-C'-C''-B-B'-B''
\ee
i.e.
\be\label{qDX4}
\qD_{X_4}=1+\qD_{\Lambda^4(Adj)}-\qD_{\Lambda^3(Adj)}-\qD_{Adj}-\qD_C-\qD_{C'}-\qD_{C''}-\qD_B-\qD_{B'}-\qD_{B''}
\ee
with
\be
\qD_{\Lambda^3(Adj)}&=&{\qD_{Adj}(u,v,w)^3\over 6}-{\qD_{Adj}(u^2,v^2,w^2)\qD_{Adj}(u,v,w)\over 2}+{\qD_{Adj}(u^3,v^3,w^3)\over 3}\nn\\
\qD_{\Lambda^4(Adj)}&=&{\qD_{Adj}(u,v,w)^4\over 24}-{\qD_{Adj}(u,v,w)^2\qD_{Adj}(u^2,v^2,w^2)\over 4}+{\qD_{Adj}(u^2,v^2,w^2)^2\over 8}+\nn\\
&+&{\qD_{Adj}(u,v,w)\qD_{Adj}(u^3,v^3,w^3)\over 3}-{\qD_{Adj}(u^4,v^4,w^4)\over 4}
\ee
and (see also \cite{MMMuniv,MkrtQDims})\footnote{$C$ is associated with $n=1$ and $k=1$ in \cite{AM1902}, while $B$ is the Cartan product of the adjoint and $Y$ representations.}
\be
\qD_B=-{\left\{{uw\over\sqrt{v}}\right\}\{w\sqrt{uv}\}\{v\sqrt{uw}\}\left\{{uv\over\sqrt{w}}\right\}\{u\sqrt{v}\}\{u\sqrt{w}\}
\{T\}\left\{{T\over\sqrt{v}}\right\}\left\{{T\over\sqrt{w}}\right\}\left\{{T\over\sqrt{u}}\right\}\over
\{\sqrt{v}\}^2\{\sqrt{w}\}^2\{\sqrt{u}\}\{u\}\left\{\sqrt{u\over v}\right\}\left\{\sqrt{u\over w}\right\}\left\{{w\over\sqrt{v}}\right\}
\left\{{v\over\sqrt{w}}\right\}}\nn\\
\qD_C=-{\{\sqrt{Tv}\}\{\sqrt{Tw}\}\{\sqrt{Tu}\}\{T\}\left\{{T\over\sqrt{v}}\right\}\left\{{T\over\sqrt{w}}\right\}
\{w\sqrt{v}\}\{v\sqrt{w}\}\left\{{vw\over u}\right\}\left\{{T\over u}\right\}\left\{{T\over v}\right\}\left\{{T\over w}\right\}\over
\left\{\sqrt{vw\over u}\right\}\left\{\sqrt{T\over v}\right\}\left\{\sqrt{T\over w}\right\}
\left\{\sqrt{u\over v}\right\}\left\{\sqrt{u\over w}\right\}\left\{{v\over\sqrt{u}}\right\}\left\{{w\over\sqrt{u}}\right\}
\{u^{3/2}\}\{\sqrt{u}\}^2\{\sqrt{v}\}\{\sqrt{w}\}}
\ee
The expression for $\qD_{X_4}$ which is obtained from these formulas is very long, and do not factorize. However, it is immediately obtained from formula (\ref{qDX4}) with all ingredients listed here. Hence, we do not write it down here. It can be found in \cite{knotebook}.

\subsection{Universal quantum dimension $\qD_{I}$\label{qDI}}

The expressions for $\qD_I$ is also very long, and do not factorize. One can look at a much simpler expression for the ordinary dimension of this representation, \cite[Eq.(3.34)]{ManeIsaevKrivMkrt} in order to get a flavor of what is the corresponding quantum dimension. The simplest way to obtain the formula for $\qD_I$ is to use the three conditions for the universal adjoint invariant that unambiguously fix the quantum dimensions of $I$, $I'$ and $I''$:
\begin{itemize}
\item
The answer for the unknot:
\be\label{unknot}
P^{[4,1]}_{Adj}(u,v,w)=1
\ee

\item The answer for the pure plethysm:
\be\label{pl}
P^{[4,0]}_{Adj}(u,v,w)={\qD_{\!_{Adj}}(u^4,v^4,w^4 )\over \qD_{\!_{Adj}}(u,v,w )}\nn\\
\ee

\item The topological invariance:
\be
P^{[4,3]}_{Adj}(u,v,w)=P^{[3,4]}_{Adj}(u,v,w)
\ee
\end{itemize}
The quantity $P^{[3,4]}_{Adj}(u,v,w)$ entering the latter condition can be obtained from \cite[Eq.(72)]{MMMuniv}, where the universal adjoint invariant of the torus knots $T[3,3k\pm 1]$ was constructed. It is of the form\footnote{The universal adjoint invariant of the torus knots $T[2,2k-1]$ constructed in \cite{MMMuniv} is of the form
\be\label{T2n}
P^{[2,n=2k-1]}_{Adj}(u,v,w) = \frac{T^{4n}}{\qD_{Adj}(u,v,w)}\cdot\Big(1-T^{-2n}\qD_{X_2}
+ T^{-2n}u^{n} \qD_{Y_2} + T^{-2n}v^{n}\qD_{Y'_2} + T^{-2n}w^{n}\qD_{Y''_2}
-T^{-n}\qD_{Adj}
\Big)
\ee
}
\be\label{T3n}
P^{[3,n=3k\pm 1]}_{Adj}(u,v,w)&=&{T^{6n}\over \qD_{Adj}(u,v,w)}\Big(1+T^{-2n}\qD_{X_3}+T^{-2n}u^{2n}\qD_{Y_3}+T^{-2n}v^{2n}\qD_{Y'_3}
+T^{-2n}w^{2n}\qD_{Y''_3}-\nn\\
&-&T^{-2n}u^n\qD_{C}-T^{-2n}v^n\qD_{C'}-T^{-2n}w^n\qD_{C''}\Big)
\ee
where
\be
\qD_{X_3}=\qD_{\Lambda^3(Adj)}-1-\qD_{X_2}-\qD_{Y_2}-\qD_{Y'_2}-\qD_{Y''_2}\nn\\
\qD_{Y_2}=\frac{\{T\}\{u\sqrt{v}w\}\{uv\sqrt{w}\}\{v\sqrt{uw}\}\{w\sqrt{uv}\}\{vw/\sqrt{u}\}}
{\{\sqrt{u}\}\{u \}\{\sqrt{v}\}\{\sqrt{w}\}\{\sqrt{u/v}\}\{\sqrt{u/w}\}}
\ee
and
\be\label{Y3}
\qD_{Y_3}= -\frac{\{uvw\}\{v\sqrt{w}\}\{w\sqrt{v}\} \{v\sqrt{uw}\}\{w\sqrt{uv}\}
\{uv\sqrt{w}\}\{uw\sqrt{v}\}\{vw/u\sqrt{u}\}\{vw\sqrt{u}\}}
{\{\sqrt{u}\}\{\sqrt{v}\}\{\sqrt{w}\} \{u\}\{u\sqrt{u}\}\{\sqrt{v}/u\}\{\sqrt{w}/u\}\{\sqrt{u/v}\}\{\sqrt{u/w}\}}
\ee
The explicit form of $I$ obtained the way described in this subsection can be found in \cite{knotebook}, while $I'$ and $I''$ are obtained from $I$ by permutations.

\subsection{Properties of the universal adjoint polynomials}

As expected, the universal adjoint polynomial (\ref{RJu}) celebrates a set of properties \cite{MMMuniv}:
\begin{itemize}
\item The special polynomial property \cite{DMMSS,IMMM,Zhu}:
\be
P^{[4,n]}_{Adj}(u=1,v=1,w)=\left(\sigma_{[1]}^{[4,n]}\right)^2
\ee
where $\sigma_{[1]}^{[4,n]}$ is the universal special polynomial in the fundamental representation. The universality is preserved at the level of special polynomials even in the fundamental representation, where one should not generally expect it. The reason is that, at $u=v=1$, the universal adjoint polynomial coincides with the HOMFLY-PT polynomial at $q=1$ upon the identification $w=A$:
\be
P^{[m,n]}_{Adj}(u=1,v=1,w)=H^{[m,n]}_{Adj}(A=w,q=1)
\ee
In particular,
\be
\sigma_{[1]}^{[4,n=2k+1]}&=&w^{3k}\left({(n-1)(n-2)(n-3)\over 6}w^6-{(n-1)(n-2)(n+1)\over 2}w^4+\right.\nn\\
&+&\left.{(n-1)(n+1)(n+2)\over 2}w^2-{(n+1)(n+2)(n+3)\over 6}\right)
\ee

\item The Alexander property of the torus knots:
\be
P^{[4,n]}_{Adj}(u,v,w)\Big|_{uvw=1}=1
\ee
The condition $uvw=1$ reduces the knot polynomial to the trivial factor, which is equal to 1 in the case of the adjoint representation and the torus knot.

\item From the Alexander property, one derives the differential expansion \cite{DE1,DE2,DE3}
\be
P^{[4,n]}_{Adj}(u,v,w)-1\ \vdots\ \{uvw\}
\ee
The remainder of this division is not universal, and depends on the concrete knot.

\item Topological invariance:
\be
P^{[4,3]}_{Adj}(u,v,w)=P^{[3,4]}_{Adj}(u,v,w)
\ee
In fact, this property is built in, along with (\ref{unknot}) and (\ref{pl}), because of the way $qD_I$'s are calculated. Moreover, one can check that the linear term in the $\hbar$-expansion of the invariant cancels, where $q=e^\hbar$:
\be
P^{[4,n]}_{Adj}(e^{\hbar\mathfrak{a}},e^{\hbar\mathfrak{b}},e^{\hbar\mathfrak{c}})=1+O(\hbar^2)
\ee
This is because the Rosso-Jones formula (\ref{RJ}) is in the topological framing.

\item Reflection invariance:
\be
P^{[4,-n]}_{Adj}(u,v,w)=P^{[4,n]}_{Adj}(u^{-1},v^{-1},w^{-1})
\ee
This property immediately follows from (\ref{RJu}) and from the invariance of the quantum dimensions in Table 2 w.r.t. to the replace $(u,v,w)\to(u^{-1},v^{-1},w^{-1})$.
\end{itemize}

\section{Conclusion}

In this paper, we constructed the universal adjoint polynomial of the knot $T[4,n]$ with odd $n$, see formula (\ref{RJu}) and Table \ref{qDu}. This adds to the previously known polynomials for the torus knots $T[3,3k\pm 1]$, formulas (\ref{T3n})-(\ref{Y3}), and $T[2n,2k-1]$, formula (\ref{T2n}). One of interesting features of the obtained results is that the knot universal adjoint polynomial does not involve the representations designated by the letters of the mathbb font in \cite{ManeIsaevKrivMkrt}: $\mathbb{X}_3$, $\mathbb{Z}_3$, $\mathbb{K}_3$, $\mathbb{L}_3$ because of (\ref{open}). These are exactly the representations that are {\it non-universally} constructed from the two representations $\hat X$ and $\tilde X$.

Known are also answers for the torus links $T[2,2k]$ and $T[3,3k]$ \cite[Eqs.(58),(79)]{MMMuniv}. Hence, in order to complete the list, one has to construct the polynomials for the links $T[4,4n]$ and $T[4,4k-2]$. In fact, constructing the universal adjoint polynomial for the link $T[4,4n]$ is quite immediate: as we explained in the Introduction, one has to use formula (\ref{RJl}) and the decomposition of the fourth power of the adjoint representation in \cite{ManeIsaevKrivMkrt}. However, it is still necessary to evaluate the quantum dimensions associated with the Casimir eigenspaces that did not emerge here in the torus knot formulas. As for the torus links $T[4,4k-2]$, in order to deal with it, first of all, one has to make the decomposition of the second degree of the Adams operation acting on the square on the adjoint representation: $\Ad_2^{\otimes 2}(Adj^{\otimes 2})$. All this requires a careful analyses, and we are planning to return to these issues elsewhere.

Another important issue related to constructing universal adjoint knot invariants of the torus knots and links is their $q,t$-deformation within the refined Chern-Simons theory. In fact, it is known \cite{KS,AM1,Mane,BM,BMM} that the refined Chern-Simons theory admits the universal formulas only for the simply laced algebras. However, though, in principle, it is known how to construct the corresponding adjoint hyperpolynomials in simple cases for concrete algebras \cite{ChE,AKMM,MMhopf}, the universal invariant for the simply laced algebras has been constructed so far only for the Hopf link $T[2,2]$ \cite{BMM}. Constructing refined universal invariants for more various cases remains a challenging problem, which deserves further studies.

\section*{Acknowledgements}

We thank A. Isaev for discussions and explanations. We are also grateful to the organizers of the Workshop ``Universal description of Lie algebras, Vogel theory, applications" in Dubna (April, 2025).
The work was partially funded within the state assignment of the Institute for Information Transmission Problems of RAS. Our work
is also partly supported by the grant of the Foundation for the Advancement of Theoretical Physics and Mathematics ``BASIS".

\section*{Appendix}

In this Appendix, we list the quantum dimensions and the eigenvalues of the second Casimir operator of all representations emerging in the Rosso-Jones formula.

\paragraph{$A$ series.}
\be
\begin{array}{rclrcl}
\qD_{[2^41^{N-8}]}&=&\displaystyle{\{Aq\}\{A/q\}^2\{A\}^2\{A/q^7\}\{A/q^2\}^2\over\{q\}^2\{q^2\}^2\{q^3\}^2\{q^4\}^2}
\hspace{1cm}& q^{\varkappa_{[2^41^{N-8}]}} &=& A^8q^{-24}\cr
\cr
\qD_{[3^42^{N-7}1^2]}&=&\displaystyle{\{A/q^6\}\{A\}^2\{A/q\}^2\{A/q^3\}\{Aq^2\}\{Aq\}\over\{q\}^3\{q^2\}^2\{q^3\}\{q^4\}}
\hspace{1cm}&q^{\varkappa_{[3^42^{N-7}1^2]}}&=&A^8q^{-16}\cr
\cr
\qD_{[4^43^{N-6}2]}&=&\displaystyle{\{A/q^5\}\{A/q^2\}\{A/q^3\}\{A\}^2\{Aq\}\{Aq^3\}\{Aq^2\}\over\{q\}^3\{q^2\}^2\{q^3\}\{q^4\}^2}
\hspace{1cm}&q^{\varkappa_{[4^43^{N-6}2]}}&=&A^8q^{-8}\cr
\cr
\qD_{[5^44^{N-5}]}&=&\displaystyle{\{A/q^3\}\{A/q^2\}\{Aq^3\}\{Aq\}\{Aq^4\}\{Aq^2\}\{A/q^4\}\{A/q\}\over\{q\}^2\{q^2\}^2\{q^3\}^2\{q^4\}^2}
\hspace{1cm}&q^{\varkappa_{[5^44^{N-5}]}}&=&A^8\cr
\cr
\qD_{[32^21^{N-7}]}&=&\displaystyle{\{A\}^2\{A/q\}^2\{A/q^3\}\{A/q^6\}\{Aq^2\}\{Aq\}\over\{q\}^3\{q^2\}^2\{q^3\}\{q^4\}^2}
\hspace{1cm}&q^{\varkappa_{[32^21^{N-7}]}}&=&A^8q^{-16}\cr
\cr
\qD_{[43^22^{N-6}1^2]}&=&\displaystyle{\{A/q^5\}\{Aq^3\}\{A/q^2\}^2\{Aq\}^2\{A\}^2\over\{q\}^4\{q^2\}^2\{q^4\}^2}
\hspace{1cm}&q^{\varkappa_{[43^22^{N-6}1^2]}}&=&A^8q^{-8}\cr
\cr
\qD_{[54^23^{N-5}2]}&=&\displaystyle{\{A/q^2\}\{A/q\}^2\{A/q^4\}\{Aq^2\}\{Aq\}^2\{Aq^4\}\over\{q\}^4\{q^2\}^2\{q^4\}^2}
\hspace{1cm}&q^{\varkappa_{[54^23^{N-5}2]}}&=&A^8\cr
\cr
\qD_{[65^24^{N-4}]}&=&\displaystyle{\{A/q^2\}\{A/q^3\}\{Aq^5\}\{A/q\}\{Aq^3\}\{Aq^2\}\{A\}^2\over\{q\}^3\{q^2\}^2\{q^3\}\{q^4\}^2}
\hspace{1cm}&q^{\varkappa_{[65^24^{N-4}]}}&=&A^8q^8\cr
\cr
\qD_{[421^{N-6}]}&=&\displaystyle{\{Aq^2\}\{A/q^3\}\{A/q^2\}\{A/q^5\}\{Aq^3\}\{Aq\}\{A\}^2\over\{q\}^3\{q^2\}^2\{q^3\}\{q^4\}^2}
\hspace{1cm}&q^{\varkappa_{[421^{N-6}]}}&=&A^8q^{-8}\cr
\cr
\qD_{[532^{N-5}1^2]}&=&\displaystyle{\{A/q^4\}\{Aq\}^2\{A/q\}^2\{Aq^2\}\{Aq^4\}\{A/q^2\}\over\{q\}^4\{q^2\}^2\{q^4\}^2}
\hspace{1cm}&q^{\varkappa_{[532^{N-5}1^2]}}&=&A^8\cr
\cr
\qD_{[643^{N-4}2]}&=&\displaystyle{\{A/q^3\}\{A\}^2\{Aq^5\}\{A/q\}^2\{Aq^2\}^2\over\{q\}^4\{q^2\}^2\{q^4\}^2}
\hspace{1cm}&q^{\varkappa_{[643^{N-4}2]}}&=&A^8q^8\cr
\cr
\qD_{[754^{N-3}]}&=&\displaystyle{\{A/q\}\{A/q^2\}\{Aq\}^2\{Aq^3\}\{A\}^2\{Aq^6\}\over\{q\}^3\{q^2\}^2\{q^3\}\{q^4\}^2}
\hspace{1cm}&q^{\varkappa_{[754^{N-3}]}}&=&A^8q^{16}\cr
\cr
\qD_{[51^{N-5}]}&=&\displaystyle{\{A/q^4\}\{A/q^3\}\{Aq^4\}\{Aq^3\}\{A/q^2\}\{Aq\}\{Aq^2\}\{A/q\}\over\{q\}^2\{q^2\}^2\{q^3\}^2\{q^4\}^2}
\hspace{1cm}&q^{\varkappa_{[51^{N-5}]}}&=&A^8\cr
\cr
\qD_{[62^{N-4}1^2]}&=&\displaystyle{\{A/q^2\}\{A/q^3\}\{Aq^5\}\{Aq^3\}\{A/q\}\{A\}^2\{Aq^2\}\over\{q\}^3\{q^2\}^2\{q^3\}\{q^4\}^2}
\hspace{1cm}&q^{\varkappa_{[62^{N-4}1^2]}}&=&A^8q^8\cr
\cr
\qD_{[73^{N-3}2]}&=&\displaystyle{\{Aq\}^2\{Aq^3\}\{A/q^2\}\{A/q\}\{Aq^6\}\{A\}^2\over\{q\}^3\{q^2\}^2\{q^3\}\{q^4\}^2}
\hspace{1cm}&q^{\varkappa_{[73^{N-3}2]}}&=&A^8q^{16}\cr
\cr
\qD_{[84^{N-2}]}&=&\displaystyle{\{A/q\}\{Aq^7\}\{Aq\}^2\{Aq^2\}^2\{A\}^2\over\{q\}^2\{q^2\}^2\{q^3\}^2\{q^4\}^2}
\hspace{1cm}&q^{\varkappa_{[84^{N-2}]}}&=&A^8q^{24}
\end{array}
\ee

\paragraph{Orthogonal series.}

\be
\hspace{-1cm}\begin{array}{rclrcl}
\qD_{[1,1,1,1]}&=&\displaystyle{ \frac{\{A q\}\{A^2 q\}\{A^2\}\{A^2/q\}}{\{q\}\{q^2\}\{q^3\}\{q^4\}}\frac{\{A^2/ q^{6}\}}{\{A/ q^{3}\}}}
\hspace{1cm}& q^{\varkappa_{[1,1,1,1]}} &=& A^8q^{-8}\cr
\cr
\qD_{[2,1,1]}&=&\displaystyle{\frac{\{Aq^2\}\{A^2q^2\}\{A^2q\}\{A^2/q\}}{\{q\}^2\{q^2\}\{q^4\}}\frac{\{A^2/q^4\}}{\{A/q^2\}}}
\hspace{1cm}&q^{\varkappa_{[2,1,1]}}&=&A^8\cr
\cr
\qD_{[3,1]}&=&\displaystyle{\frac{\{Aq^3\}\{A^2 q^3\}\{A^2 q\}\{A^2\}}{\{q\}^2\{q^2\}\{q^4\}} \frac{\{A^2/q^2\}}{\{A/q\}}}
\hspace{1cm}&q^{\varkappa_{[3,1]}}&=&A^8q^{8}\cr
\cr
\qD_{[4]}&=&\displaystyle{\frac{\{Aq^4\}\{A^2q^3\}\{A^2q^2\}\{A^2q\}}{\{q\}\{q^2\}\{q^3\}\{q^4\}} \frac{\{A^2\}}{\{A\}}}
\hspace{1cm}&q^{\varkappa_{[4]}}&=&A^8q^{16}\cr
\cr
\qD_{[1^8]}&=&\displaystyle{\frac{\{Aq\}\{A^2q\}\{A^2\}\{A^2/q\}\{A^2/q^2\}\{A^2/q^3\}\{A^2/q^4\}\{A^2/q^5\}}{\{q\}\{q^2\}\{q^3\}\{q^4\}\{q^5\}\{q^6\}\{q^7\}\{q^8\}} \frac{\{A^2/q^{14}\}}{\{A/q^{7}\}}}
\hspace{1cm}&q^{\varkappa_{[1^8]}}&=&A^{16}q^{-48}\cr
\cr
\qD_{[21^6]}&=&\displaystyle{\frac{\{Aq^2\}\{A^2q^2\}\{A^2q\}\{A^2\}\{A^2/q\}\{A^2/q^2\}\{A^2/q^3\}\{A^2/q^5\}}{\{q\}^2\{q^2\}\{q^3\}\{q^4\}\{q^5\}\{q^6\}\{q^8\}} \frac{\{A^2/q^{12}\}}{\{A/q^{6}\}}}
\hspace{1cm}&q^{\varkappa_{[21^6]}}&=&A^{16}q^{-32}\cr
\cr
\qD_{[2^4]}&=&\displaystyle{\frac{\{Aq\}\{Aq^2\}\{A^2q^3\}\{A^2q^2\}\{A^2q\}^2\{A^2\}\{A^2/q^5\}}{\{q\}\{q^2\}^2\{q^3\}^2\{q^4\}^2\{q^5\}} \frac{\{A^2/q^{6}\}\{A^2/q^{4}\}}{\{A/q^{3}\}\{A/q^{2}\}}}
\hspace{1cm}&q^{\varkappa_{[2^4]}}&=&A^{16}q^{-8}\cr
\cr
\qD_{[31^5]}&=&\displaystyle{\frac{\{Aq^3\}\{A^2q^3\}\{A^2q^2\}\{A^2q\}\{A^2\}\{A^2/q\}\{A^2/q^3\}\{A^2/q^4\}}{\{q\}^2\{q^2\}^2\{q^3\}\{q^4\}\{q^5\}\{q^8\}} \frac{\{A^2/q^{10}\}}{\{A/q^{5}\}}}
\hspace{1cm}&q^{\varkappa_{[31^5]}}&=&A^{16}q^{-16}\cr
\cr
\qD_{[3,2,2,1]}&=&\displaystyle{\frac{\{Aq^3\}\{Aq\}\{A^2q^4\}\{A^2q^3\}\{A^2q\}^2\{A^2/q\}\{A^2/q^4\}}{\{q\}^3\{q^2\}\{q^3\}\{q^4\}^2\{q^6\}} \frac{\{A^2/q^{6}\}\{A^2/q^2\}}{\{A/q^{3}\}\{A/q\}} }
\hspace{1cm}&q^{\varkappa_{[3,2,2,1]}}&=&A^{16}\cr
\cr
\qD_{[3,3,2]}&=&\displaystyle{\frac{\{Aq^2\}\{Aq^3\}\{A^2q^5\}\{A^2q^3\}\{A^2q^2\}\{A^2\}\{A^2/q\}\{A^2/q^3\}}{\{q\}^2\{q^2\}^2\{q^3\}\{q^4\}^2\{q^5\}} \frac{\{A^2/q^{2}\}\{A^2/q^4\}}{\{A/q\}\{A/q^2\}} }
\hspace{1cm}&q^{\varkappa_{[3,3,2]}}&=&A^{16}q^{8}\cr
\cr
\qD_{[4,1^4]}&=&\displaystyle{\frac{\{Aq^4\}\{A^2q^4\}\{A^2q^3\}\{A^2q^2\}\{A^2q\}\{A^2/q\}\{A^2/q^2\}\{A^2/q^3\}}{\{q\}^2\{q^2\}^2\{q^3\}^2\{q^4\}\{q^8\}} \frac{\{A^2/q^{8}\}}{\{A/q^4\}} }
\hspace{1cm}&q^{\varkappa_{[4,1^4]}}&=&A^{16}\cr
\cr
\qD_{[4,2,1,1]}&=&\displaystyle{\frac{\{Aq^4\}\{Aq\}\{A^2q^5\}\{A^2q^3\}\{A^2q^2\}\{A^2\}\{A^2/q\}\{A^2/q^3\}}{\{q\}^3\{q^2\}^2\{q^4\}^2\{q^7\}} \frac{\{A^2\}\{A^2/q^6\}}{\{A\}\{A/q^3\}} }
\hspace{1cm}&q^{\varkappa_{[4,2,1,1]}}&=&A^{16}q^{8}\cr
\cr
\qD_{[4,3,1]}&=&\displaystyle{\frac{\{Aq^4\}\{Aq^2\}\{A^2q^6\}\{A^2q^3\}\{A^2q\}^2\{A^2/q\}\{A^2/q^2\}}{\{q\}^3\{q^2\}\{q^3\}\{q^4\}^2\{q^6\}} \frac{\{A^2\}\{A^2/q^4\}}{\{A\}\{A/q^2\}} }
\hspace{1cm}&q^{\varkappa_{[4,3,1]}}&=&A^{16}q^{16}\cr
\cr
\qD_{[4,4]}&=&\displaystyle{\frac{\{Aq^4\}\{Aq^3\}\{A\}\{A^2q^7\}\{A^2q^2\}\{A^2q\}^2\{A^2/q\}}{\{q\}\{q^2\}^2\{q^3\}^2\{q^4\}^2\{q^5\}} \frac{\{A^2\}^2\{A^2/q^2\}}{\{A\}^2\{A/q\}} }
\hspace{1cm}&q^{\varkappa_{[4,4]}}&=&A^{16}q^{24}\cr
\cr
\end{array}
\ee


\begin{thebibliography}{12}

\bibitem{Vogel95} P. Vogel, {\sl Algebraic structures on modules of diagrams}, Preprint (1995), available at \url{https://webusers.imj-prg.fr/~pierre.vogel/diagrams.pdf}, J. Pure Appl. Algebra, {\bf 215} (2011) 1292-1339

\bibitem{Vogel99}  P. Vogel, {\sl The Universal Lie algebra}, Preprint (1999), available at \url{https://webusers.imj-prg.fr/~pierre.vogel/grenoble-99b.pdf}
    

\bibitem{KMS} D.~Khudoteplov, A.~Morozov, A.~Sleptsov,
{\sl Can Yang-Baxter imply Lie algebra?},
arXiv:2503.13437

\bibitem{MkrtVes12} R.L. Mkrtchyan, A.P. Veselov, 
JHEP {\bf 08} (2012) 153, arXiv:1203.0766

\bibitem{Mkrt13} R.L. Mkrtchyan, 
JHEP {\bf 09} (2013) 54, arXiv:1302.1507

\bibitem{KreflMkrt} D. Krefl, R. Mkrtchyan, 
JHEP {\bf 10} (2015) 45, arXiv:1506.03907

\bibitem{M2} R.L.~Mkrtchyan,
J. Geom. Phys. \textbf{129} (2018) 186-191,
arXiv:1709.03261

\bibitem{Westbury03} B. Westbury, 
Proceedings of the Tenth Oporto Meeting on Geometry, Topology and Physics, {\bf 18} (2001) suppl. 2003, pp. 49-82


\bibitem{MkrtQDims} R.L. Mkrtchyan, 
Nucl. Phys. {\bf B921} (2017) 236-249, arXiv:1610.09910

\bibitem{LandMan06} J.M. Landsberg, L. Manivel, 
Adv. Math. {\bf 201} (2006) 379-407

\bibitem{MkrtSergVes} R.L. Mkrtchyan, A.N. Sergeev, A.P. Veselov, 
J. Math. Phys. {\bf 53} (2012) 102106, arXiv:1105.0115

\bibitem{ManeIsaevKrivMkrt} M.Y. Avetisyan, A.P. Isaev, S.O. Krivonos, R.L. Mkrtchyan, 
Russian Journal of Mathematical Physics, {\bf 31} (2024) 379-388, arXiv:2311.05358

\bibitem{IsaevProv} A.P. Isaev, A.A. Provorov, 
Theor.Math.Phys. {\bf 206} (2021) 3-22, arXiv:2012.00746

\bibitem{IsaevKriv} A.P. Isaev, S.O. Krivonos, 
 J. Math. Phys. {\bf 62} (2021) 083503, arXiv:2102.08258

\bibitem{IsaevKrivProv} A.P. Isaev, S.O. Krivonos, A.A. Provorov, 
    Int. J. Mod. Phys. {\bf A38} (2023) 235003, arXiv:2212.14761

\bibitem{KhM} H.M.~Khudaverdian, R.L.~Mkrtchyan,
Lett. Math. Phys. \textbf{107} (2017) 1491-1514,
arXiv:1602.00337

    \bibitem{MMMuniv} A. Mironov, R. Mkrtchyan, A. Morozov, 
    JHEP {\bf 02} (2016) 078, arXiv:1510.05884

\bibitem{MMuniv} A. Mironov, A. Morozov, 
Phys. Lett. {\bf B755} (2016) 47-57, arXiv:1511.09077

    \bibitem{KLS} D.~Khudoteplov, E.~Lanina, A.~Sleptsov,
{\sl Construction of Lie algebra weight system kernel via Vogel algebra},
arXiv:2411.14417

\bibitem{BMM} L.~Bishler, A.~Mironov, A.~Morozov,
{\sl Macdonald deformation of Vogel's universality and link hyperpolynomials},
arXiv:2505.16569

\bibitem{Vivek} Vivek Kumar Singh et al., to appear

\bibitem{RJ} M. Rosso, V.F.R. Jones, J. Knot Theory Ramifications, \textbf{2} (1993) 97-112

\bibitem{China1} X.-S. Lin, H. Zheng, Trans.Amer.Math.Soc. \textbf{362} (2010) 1-18 math/0601267

\bibitem{Stevan} S. Stevan, Annales Henri Poincare {\bf 11} (2010) 1201-1224, arXiv:1003.2861

\bibitem{China2} Lin Chen, Qingtao Chen, 
Pacific Journal of Mathematics {\bf 257} (2012) 267-318, arXiv:1007.1656

\bibitem{modRT} A.~Mironov, A.~Morozov, An.~Morozov,
JHEP \textbf{03} (2012) 034,
arXiv:1112.2654

\bibitem{SF} \url{https://www.symmetricfunctions.com/}

\bibitem{M} F.D. Murnaghan, {\sl The Theory of Group Representations}, (The Johns Hopkins Press,
Baltimore, 1938)

\bibitem{L} D.E. Littlewood, {\sl The Theory of Group Characters and Matrix Representations of Groups}, (The Clarendon Press, Oxford, 1950

\bibitem{BC} A.B.~Balantekin, P.~Cassak,
J. Math. Phys. \textbf{43} (2002) 604-620,
arXiv:hep-th/0108130

\bibitem{Mac}   I.G. Macdonald,
  \textit{Symmetric functions and Hall polynomials},
  Oxford University Press, 1995

\bibitem{ChE} I.~Cherednik, R.~Elliot,
{\sl Refined composite invariants of torus knots via DAHA},
arXiv:1503.01441

\bibitem{R81} R.~L.~Mkrtchian,
Phys. Lett. \textbf{B105} (1981) 174-176

\bibitem{Tsv} P. Cvitanovic, {\sl Group Theory}, Princeton University Press, Princeton, NJ, 2004

\bibitem{MV10} R.L.~Mkrtchyan, A.P.~Veselov,
J. Math. Phys. \textbf{52} (2011) 083514,
arXiv:1011.0151

\bibitem{BFM} V.~Bouchard, B.~Florea, M.~Marino,
JHEP \textbf{12} (2004) 035,
hep-th/0405083

\bibitem{AM1902} M.~Y.~Avetisyan, R.~L.~Mkrtchyan,
{\sl On universal quantum dimensions of certain two-parameter series of representations},
arXiv:1909.02076

\bibitem{knotebook} \url{http://wwwth.itep.ru/knotebook/knotebook/HOMFLY/universal.htm}

\bibitem{DMMSS} P.~Dunin-Barkowski, A.~Mironov, A.~Morozov, A.~Sleptsov, A.~Smirnov,
JHEP \textbf{03} (2013) 021,
arXiv:1106.4305

\bibitem{IMMM} H.~Itoyama, A.~Mironov, A.~Morozov, A.~Morozov,
JHEP \textbf{07} (2012) 131,
arXiv:1203.5978

\bibitem{Zhu} S.~Zhu,
JHEP \textbf{10} (2013) 229,
arXiv:1206.5886

\bibitem{DE1}  N.M. Dunfield, S. Gukov, J. Rasmussen, Experimental Math. {\bf 15} (2006) 129-159, math/0505662

\bibitem{DE2} H.~Itoyama, A.~Mironov, A.~Morozov and A.~Morozov,
JHEP \textbf{07} (2012) 131,
arXiv:1203.5978

\bibitem{DE3} A.~Mironov, A.~Morozov and A.~Morozov,
AIP Conf. Proc. \textbf{1562} (2013) 123-155,
arXiv:1306.3197

\bibitem{KS} D. Krefl, A. Schwarz, 
Journal of Geometry and Physics, {\bf 74} (2013) 119-129, arXiv:1304.7873

\bibitem{AM1} M.Y.~Avetisyan, R.L.~Mkrtchyan,
JHEP \textbf{10} (2021) 033,
arXiv:2107.08679

\bibitem{Mane} M. Avetisyan, {\sl Vogel's Universality and its Applications}, arXiv:2207.04302

\bibitem{BM} L.~Bishler, A.~Mironov,
Phys. Lett. \textbf{B867} (2025) 139596,
arXiv:2504.13831

\bibitem{AKMM} H.~Awata, H.~Kanno, A.~Mironov, A.~Morozov,
Nucl. Phys. \textbf{B949} (2019) 114816,
arXiv:1905.00208

\bibitem{MMhopf} A.~Mironov, A.~Morozov,
Nucl. Phys. \textbf{B960} (2020) 115191,
arXiv:2003.07836

\end{thebibliography}
\end{document}